\documentclass[authoryear,preprint]{elsarticle}
\usepackage{natbib}

\usepackage{aas_macros}
\usepackage{amssymb}

\usepackage{graphicx}
\usepackage{url}

\newcommand{\um}{\mbox{$\mu\mathrm{m}$}}
\newcommand{\uJy}{\mbox{$\mu\mathrm{Jy}$}}
\newcommand{\MJy}{\mbox{$\mathrm{MJy}$}}
\newcommand{\Jy}{\mbox{$\mathrm{Jy}$}}
\newcommand{\m}{\mbox{m}}
\newcommand{\km}{\mbox{km}}
\newcommand{\AU}{\mbox{$\mathrm{AU}$}}
\newcommand{\pc}{\mbox{$\mathrm{pc}$}}
\newcommand{\K}{\mbox{$\mathrm{K}$}}
\newcommand{\uK}{\mbox{$\mu\mathrm{K}$}}
\newcommand{\perK}{\mbox{$\mathrm{K}^{-1}$}}
\newcommand{\persr}{\mbox{$\mathrm{sr}^{-1}$}}
\newcommand{\persqarcsec}{\mbox{$\mathrm{arcsec}^{-2}$}}
\newcommand{\persec}{\mbox{$\mathrm{s}^{-1}$}}
\newcommand{\Tsys}{\mbox{$T_\mathrm{sys}$}}
\newcommand{\Teff}{\mbox{$T_\mathrm{eff}$}}
\newcommand{\Rsun}{\mbox{$R_\mathrm{Sun}$}}
\newcommand{\Rearth}{\mbox{$R_\mathrm{Earth}$}}

\begin{document}

\title{Habitable Planet Detection and Characterization with Far Infrared Coherent Interferometry}

\author{James P. Lloyd}
\ead{jpl@astro.cornell.edu}
\address{Department of Astronomy, Cornell University, Ithaca NY}

\begin{abstract}

The characterization of extrasolar earth-like atmospheres for biosignatures remains one of the most
compelling and elusive challenges in astronomy.  Coronagraphy, nulling interferometry and free-flying occulters have been advanced as potential techniques to accomplish this gaol.  In this paper, a complementary approach, coherent interferometry in the far infrared is considered.  For an interferometer operating close to the quantum noise limit, a collecting area of $\sim$ 1000 m$^{2}$ and baselines of 200 m are sufficient to detect an earth-like planet.  The high spectral resolution achievable with coherent detection further enables unambiguous molecular inventory of an atmosphere and retrieval of atmospheric temperature-pressure-composition profiles.  The far-infrared is rich in molecular features, particularly transitions of the key biosignature molecules H$_2$O and O$_3$.   The level of detail that can be obtained on atmospheres is such that the goals of detection and detailed characterization of biosignatures can be accomplished by the same mission.  Hitherto, however, the majority of modeling efforts concerning the extrasolar planet atmospheres has been limited to the visible and thermal infrared regimes considered for coronagraphs and nulling interferometry.  It is therefore worth seriously investigating the feasibility of such an architecture for a possible mission, and considering biosignatures that might be available in the far-infrared.

\end{abstract}


\maketitle
%
%
%

\section{Introduction}

The allure of extrasolar planets has inspired an armada of detection techniques, including recent remarkable success stories that have brought forth discoveries of hundreds of planets.  Planets orbiting other stars have now been directly imaged \citep{Kalas:2008fr,Marois:2008zr,Lafreniere:2010fj,Lagrange:2009kx}, but in some respects have have proven to be more enigmatic than illuminating, leading to controversy as to their formation mechanism \citep{Boley:2009qy} or even whether they are ``planets'' \citep{Kratter:2010uq}.  Indirect approaches have been the most successful, not only in terms of sheer count of known planets, but remarkably in revelation of physical properties of those {\em indirectly detected} planets.   This proliferation of exoplanet particulars has been something of a surprise in the context of prior presumptions that the detailed characterization of extrasolar planets would necessarily await a space-based coronagraph, occulter or nulling interferometer.  In 1835, August Comte predicted that we should never understand the composition of celestial bodies only by the light from those objects.  Akin to the contrast of that statement with modern knowledge of stars obtained through telescopic observations, it is remarkable how the planetary science of extrasolar planets has thrived even in the absence of detailed results from targeted missions.

The combination of mass and radius revealed by the {\em joint} measurements of Doppler radial velocity and transit depth measures density.  When combined with an equation of state for cold degenerate matter, density infers bulk composition in the same way composition of the solar system planets was determined \citep{Zapolsky:1969fk}.  A menagerie, now exceeding 100 planets, has been characterized this way with diverse properties of individual planets ranging from: near-cosmic abundance cold-degenerate objects very comparable to Jupiter (e.g., \citet{Cameron:2007uq}); highly enriched in heavy elements implying large rocky cores (e.g. \citet{Sato:2005fj}); similar to Neptune and Uranus (e.g. \citet{Gillon:2007vn,Borucki:2010rt,Bakos:2010ys}), and large, low density planets \citep{Charbonneau:2000rr,Mandushev:2007dq} only explained by significant internal heating \citep{Bodenheimer:2001kl,Winn:2005dp,Jackson:2008tg,Batygin:2010hc} or revisions to structure and evolution models \citep{Arras:2006eu,Burrows:2007ai,Youdin:2010oq,Arras:2010qe}.  Sufficiently precise photometry to detect an Earth-radius planet is only possible from space, and in pursuit of this goal the CoRoT and Kepler missions have been launched. Kepler has discovered  a rocky planet only 1.4 times the size of Earth \citep{Batalha:2011lr} in a 0.84 day period orbit, and has a bounty of planets with measured radii \citep{Borucki:2011vn}.  By concentrating on surveying the smallest stars we can detect transits by the smallest planets \citep{Nutzman:2008vn}.  The first such discovery is GJ 1214b, a super-earth with evidence for a likely atmosphere of hydrogen/helium or possibly water \citep{Charbonneau:2009qy}.  

Precision Doppler measurements of 166 stars now show that rocky planets are common, and predict 23\% of solar-type stars have a planet of mass 0.5-2 M$_{\mathrm{Earth}}$ in orbits less than 50 days \citep{Howard:2010fr}.  Kepler is already finding a bounty of transits that will unfortunately be an overwhelming challenge to authenticate with ground-based Doppler measurements \citep{Borucki:2011vn}. In the absence of a Doppler mass measurement, not only is the planetary density (and therefore composition) unknown, but there remains an admitted possibility of any number of astrophysical false positives \citep{Brown:2003ul,Charbonneau:2003qf,Charbonneau:2004gf,Torres:2004ly,ODonovan:2006mz,Evans:2010ve}.   In some cases confirmation by dynamical analysis of multi-planet systems \citep{Holman:2010rt,Lissauer:2011ys} or meticulous elimination of false positives \citep{Torres:2011pd} can assuage the residual skepticism. In many cases, however, even though a photometric signal can be detected, the sources may simply prove too faint to convincingly eliminate the qualification of "candidate" to the transiting planet (e.g. \citet{Sahu:2006lq}).

At the other extreme, for the most favorable case of large, massive planets transiting bright, nearby stars, observations of sufficient fidelity have elicited understanding of physical processes far beyond characterization with a single parameter such as density.  In these cases, it is possible to disentangle the planet and starlight temporally, if not spatially.  Since the transit depth measures the projected optical size of the planet, differential (in vs out of transit) spectroscopy has shown absorption from species including sodium \citep{Charbonneau:2002cr}, HCN and C$_2$H$_2$ \citep{Shabram:2011lr}, and atmospheric haze \citep{Pont:2008uq}.

Transiting planetary systems  additionally show an eclipse of the planet by the star, and similarly differential observations can measure thermal emission \citep{Charbonneau:2005wd} and reflected light \citep{Borucki:2009gf} of the planet.  These emission spectra have shown signatures including water \citep{Tinetti:2007fk,Burrows:2007yq}, methane and carbon dioxide \citep{Swain:2009rt},  silicate clouds \citep{Richardson:2007qy}, stratospheric temperature inversions \citep{Knutson:2008fk,Knutson:2009kx}, and thermochemical disequilibrium \citep{Stevenson:2010yq}.

Since hot Jupiters are in general tidally locked, one face is strongly irradiated and one face not, resulting in longitudinal temperature variations that modulate the observed  flux as the planet rotates \citep{Harrington:2006nx}.   These brightness variations can indicate hydrodynamical redistribution of the incident energy \citep{Knutson:2007ve} and in the case of eccentric planets, dramatic heating and cooling \citep{Laughlin:2009ul}.

In some cases the proximity of the planet to the star results in the heating of a hydrodynamically escaping thermosphere. This thermosphere can fill the Roche lobe, and be optically thick in resonance lines resulting in transit depths enhanced by a factor of 10.  Resonance line absorption from a thermosphere has been observed in Ly $\alpha$ \citep{Vidal-Madjar:2003mz}, oxygen and carbon \citep{Vidal-Madjar:2004ly}, neutral sodium, tin, and manganese, and singly ionized ytterbium, scandium, manganese, aluminum, vanadium, and magnesium \citep{Fossati:2010ys} and triply ionized silicon \citep{Schlawin:2010qf}.

Transiting planets have revealed fossil evidence of their dynamical history through the spin-orbit alignment with the star, which can be measured through the Rossiter-McLaughlin effect.  A range of alignments is observed, which informs theories of the migration history \citep{Fabrycky:2009lq} of hot Jupiters, and the tidal interactions with the star \citep{Winn:2010bh}.  Multiple planet systems close to resonance seem to be an inevitable consequence of migration processes, and muti-body gravitational interactions will lead to transit timing variations that can be sensitive to extremely low mass additional planets \citep{Agol:2005dq,Holman:2005rr}.

\section{Towards Other Earths}

Extrapolation of this influx of progress leads to the expectation that the discovery and characterization of earth-like planets is the inevitable consequence of the current momentum of the field.  Indeed, based purely on a bootstrap analysis of the rate of discovery, \citet{Arbesman:2010cr} predict that the first ``earth-like'' planet will be announced by May 2011.  The surprising breadth and depth of the discoveries enabled by novel approaches to transiting planet observations has prompted extensive exploration of the limits of what can be accomplished in characterization of transiting exoplanets, including dedicated space telescopes for spectroscopic transit observations \citep{Swain:2010lr}, and all sky transit searches to completely catalog the nearest transiting planets \citep{Deming:2009fk,Catala:2009qy}.  

It is tempting to assume that successes of transiting hot jupiter planets could be extrapolated to smaller and possibly habitable planets in the same way atmospheric measurements of hot jupiter planets have been obtained, if only a sufficiently high SNR could be obtained.  However, as the collecting area or integration time in combined light is proportional to the SNR (or equivalently contrast) squared, the requirements rapidly become prohibitive.  The current regime of transiting hot jupiters is operating at contrasts around $10^3$.  To push this to the contrast limits of an earth-twin, SNR $\sim10^3$ times larger is be required.  Assuming simple photon noise, this could be met with $10^6$ times the collecting area or integration time.  Without separating out the starlight, pushing from where we are now to the requisite contrast increases the demands so square meters become square kilometers, seconds of integration become days of integration, or hours of integration become centuries.   While combined light spectroscopy is feasible for the modest contrast of $10^3$ in the cases of hot Jupiters, and will surely reveal remarkable discoveries of the properties of super-earths, study of the earth-like terrestrial planets that are expected to be abundant in the Universe at the requisite contrasts of $10^6$--$10^9$ push this technique beyond reasonable limits.

A large fraction of the remarkable discoveries above rely on observations from the 0.85 meter Spitzer Space Telescope, which takes advantage of the inherently stable platform of a space observatory and the exquisite sensitivity achievable in the infrared from cold space telescopes.  Spitzer is now only operational in two wavelength channels, but the 6.5 meter James Webb Space Telescope (JWST), when launched, will dramatically improve upon results from Spitzer, having 50 times the collecting area, and improved spectroscopic capabilities.  The signal to noise ratio for bright object observations  improves with the square root of collecting area, so JWST will provide a welcome, but moderate improvement in SNR of a factor of approximately 7.  Particularly when combined with improved spectroscopic capabilities, there will of course be a steady stream of results akin to those above on hot Jupiters.  The task of atmospheric characterization of earth-like planets, particularly those discovered by Kepler, which is only observing distant stars in a narrow field, will be well beyond the sensitivity of JWST.  Calculations of the absorption spectra of earth-like planets observed with JWST \citep{Kaltenegger:2009fr} demonstrate that molecular absorption features are generally detected with SNR less than one per transit for nearby stars.   Although the transit depth for an earth-like planet is challengingly small, the atmospheric absorption depth is partially offset by the lower gravity of a small planet resulting in a larger atmospheric scale height \citep{Beckwith:2008zr}.  Co-adding many transits can improve the SNR, but the transits for a planet in the habitable zone of a G star occur but once per year, so the characterization of habitable zone planets with JWST will be restricted to planets in the habitable zone of late M stars that will undergo many transits over the JWST mission lifetime.  No such transiting planet is known, although efforts are underway to identify such a planet by specifically targeting low mass stars for planet searches.   Such planets have been traditionally considered as poor sites for potential life as the tidal locking of the planet may result in cold-trapping of the atmosphere on the night side \citep{Kasting:1993kx}.  There are possible climates that would avert such a catastrophe \citep{Pierrehumbert:2011uq}, which may show detectable signatures \citep{Pierrehumbert:2010lr}.   There are also counter-arguments to most objections to the habitability of planets orbiting M stars \citep{Lammer:2007ys}, but in general these scenarios diverge wildly from "earth-like" in all respects but the radiative equilibrium temperature of the planet being near 300K.   There is an opportunity in the short term to characterize the terrestrial planets of M stars \citep{Lunine:2008lr}, but this opportunity is only an brief prelude to the challenge of answering the question of how rare are planets like the Earth and the life they bear.

\section{Identification of Earth-like Planets}

Despite the flood of progress in extrasolar planets with modest facilities, a sober assessment places ``earth-twin'' characterization well into the indefinite future.  We will continue to identify through indirect techniques generic signposts to earth-like planets: objects which we know exist and infer that liquid water, or even life, {\em might} exist, but when and how will we catalog a place where it {\em does}?

The planets in our own solar system remain the only ones of which we have the detailed knowledge necessary to make definitive statements concerning habitability and even then great uncertainty remains.  There might be occupied or unoccupied exotic niches for life on worlds such as Mars, Europa, Enceladus, Titan, or even Earth.  It is of note that these particular sites of serious scientific interest are only afforded that status by secure results indicating liquid activity from planetary missions \citep{Reynolds:1983fj,Lunine:1983pd,Porco:2006lr}.  Spectroscopic evidence \citep{Houck:1973kx} of water on Mars  did not give way to the belief that Mars had a past with liquid water amenable to life until unambiguous proof was at hand \citep{Squyres:2004qf}.  The recent history of planetary science (see \citet{Burns:2010fk} for a review) is dominated by discoveries made through exploration with spacecraft, but in the four hundred years since Galileo remarkable progress was also made through remote sensing.   The daunting scales and physical constraints of interstellar travel must cede the prospect of probes to extrasolar planets to the realm of science fiction for the indefinite future.  While characterization of habitable planets through remote sensing remains in the future, few would consider it to be indefinitely so.

\begin{figure}[htbp]
\begin{center}
\includegraphics[width=\columnwidth]{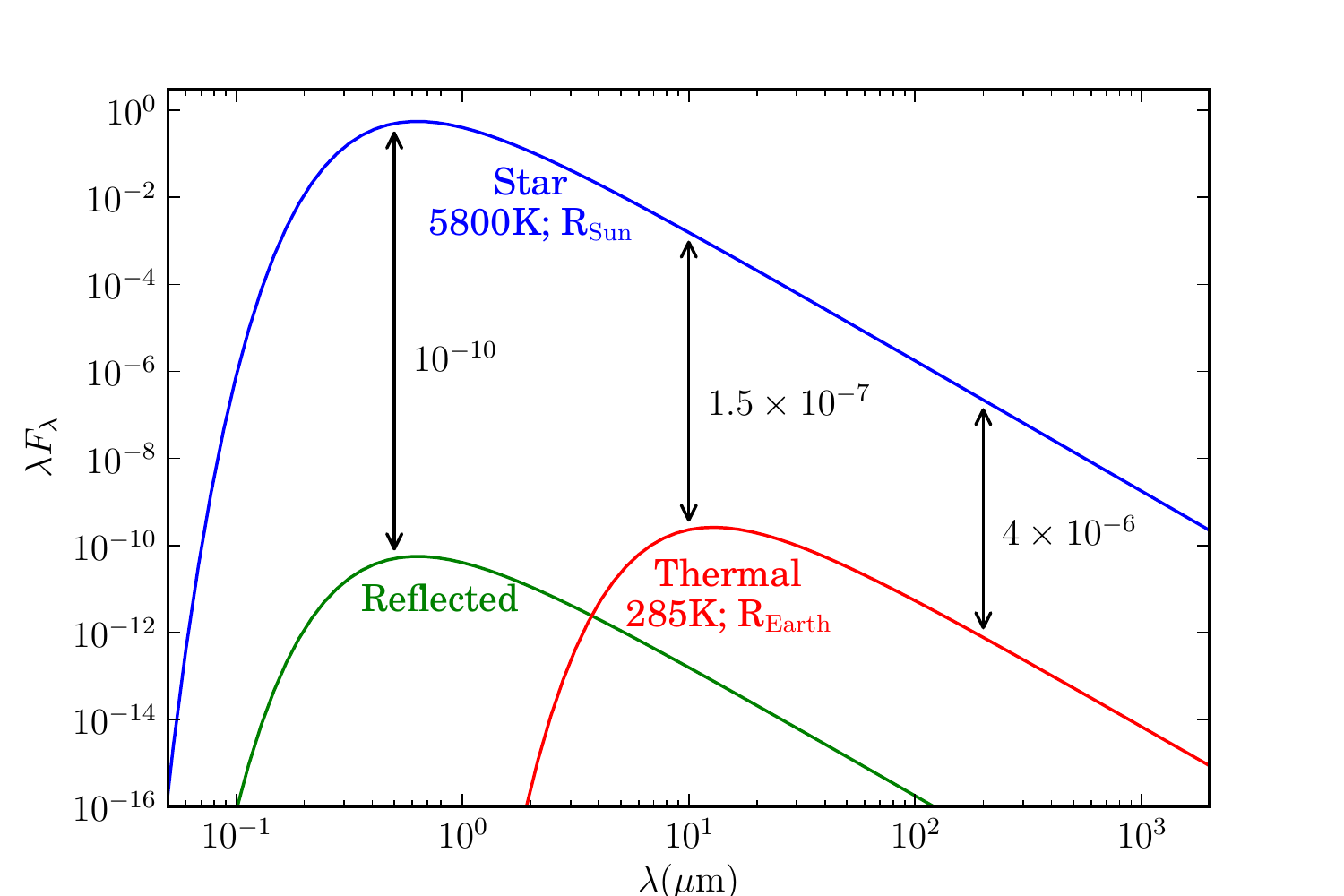}
\caption{Model stellar and planetary thermal and reflected fluxes.  The star is modelled as a $\Teff=5800 \K$ blackbody of $R=1 \Rsun$, and the planet as a $\Teff=285 \K$ blackbody of $R=1 \Rearth$.  The reflected component is modeled as $10^{-10}$ of the stellar flux, representative of an average albedo and phase angle for an earth-like planet at $a=1 \AU$.  At the wavelength region TPF-I is planned to operate near the Ozone feature, $10~\um$, the contrast is $1.5\times10^{-7}$, and on the Rayleigh-Jeans tail, the contrast asymptotes to $T_{\mathrm{p}}/T_\star  (R_{\mathrm{p}}/R_\star)^2 \sim 4\times10^{-6}$. }
\label{fig:visvsir}
\end{center}
\end{figure}

Concepts for characterizing planets have generally focussed on separation of the planet light from the photon noise of the much brighter star-light.  Beginning more than thirty years ago concepts for planet detection from space in the visible \citep{Bonneau:1975ly,Angel:1986fk} and infrared \citep{Bracewell:1978lr} have become highly developed, following a recommended investment of \$200M by the 2000 US decadal survey in technology for the Terrestrial Planet Finder (TPF) mission, originally conceived to be a nulling interferometer operating in the infrared.  The infrared affords favorable contrast ratios compared to the visible (see Figure~\ref{fig:visvsir}), but dictates a cryogenic and likely formation flying instrument.  Following a proliferation of concepts for coronagraphs, the direct imaging of planets in the visible has become a viable competitor.  Detailed designs have been completed for a the Terrestrial Planet Finder (TPF) mission with a coronagraph (TPF-C) \citep{Traub:2006gf}), interferometer (TPF-I/Darwin) \citep{Lawson:2007ve} and free-flying occulter (TPF-O) \citep{Cash:2007ul}, all of which were at one time proposed for launch early this decade.   Arguments have been made that the combination of visible, infrared, and mass measurements with an astrometric interferometer are necessary to completely characterize a potentially habitable planet \citep{Beichman:2007fk}.  Programmatic decisions have pushed any possible launch of any of these missions into the indefinite future, with the most optimistic scenario being a TPF-like mission entering development after the 2020 US decadal survey.  

In the absence of noise from the star, the collecting area required to detect a nearby earth in broadband blackbody thermal emission is not necessarily large \citep{Bally:1994fk},

\begin{eqnarray}
A _{e}& = & 8~\mathrm{m}^2 \left( \frac{B}{ 1 \mathrm{~MJy~sr^{-2}}} \right) ^{0.5} 
        \left( \frac{\lambda}{{100~\mu\mathrm{m}}} \right)^{1.5} \nonumber\\
    &&
       \left( \frac{ T_p}{300~\mathrm{K}} \right)^{-1} 
       \left( \frac{d}{10~\mathrm{pc}} \right)^2 
           \left( \frac{r_p}{10^9~\mathrm{cm}} \right)^{-2} 
       \left( \frac{t}{10^4~\mathrm{s}} \right)^{-0.5} 
      \left( \frac {\Delta\nu}{100~\mathrm{GHz}}\right)^{-0.5}
\end{eqnarray}

where $A_{e}$ is the effective area required to detect the source with SNR=1, $B$ is the background emission, $\lambda$ is the wavelength of observation, $T_p$ is the blackbody temperature of the planet, $d$ is the distance to the planet, $r_p$ is the size of the planet, $t$ is the integration time, and $\Delta\nu$ is the bandwidth of the integration.  Indeed it is this fact that makes TPF-I an attractive mission.  As long as the challenge of suppressing the starlight can be met, the collecting area requirements are modest.  Unfortunately the nulling and resolution requirements inevitably lead to a minimum of four formation-flying spacecraft, and maintaining low background requires cryogenic optics, leading to a complex and expensive mission.

 In the case of TPF, the mission requirements are specified as sufficient to detect the bands of primary atmospheric constituents relevant to life, such as H$_2$O, O$_2$, O$_3$, CH$_4$ and CO$_2$.  TPF will only achieve the necessary sensitivity at modest spectral resolution, $\Delta \lambda/\lambda \lesssim 100$.  Interpreting such a spectrum will rely on a number of assumptions and model calculations.  There are a wide range of possible geological, atmospheric and climate states that are known to have occurred on Earth, which will result in a variety of spectral features \citep{Kaltenegger:2007qy}.   Any hypothetical extrasolar terrestrial planet could exist in geological, geochemical, atmospheric or climate state that are similar, or completely unlike any of these.  Although many common species are well characterized, it is always possible trace species may produce unexpected features, for example recently discovered absorption by isotopologues of CO$_2$ in the atmosphere of Mars that produce an absorption feature at 3.3 $\mu$m, near important biomarker features such as CH$_4$ \citep{Villanueva:2008yq}. In addition to the atmospheric absorption, surface geology, texture, continents, oceans and vegetation will introduce absorption bands \citep{Seager:2005uq,Hamdani:2006fj}.  An essentially infinite number of degrees of freedom exist in the geology, geochemistry, meteorology, atmospheric chemistry, geography, and biology that might exist on other worlds.   All reflectance, absorption and emission characteristics on the Earth-facing side of a given planet will be integrated into a single spatial resolution element, and recorded as a moderate resolution spectrum at finite signal to noise ratio.  The inherent ambiguity and model-dependance in moderate resolution spectra may forever thwart this approach from meeting the standard of extraordinary evidence that will be demanded of the extraordinary claim of life on another planet.

In the short term, in spite of the extraordinary promise of the discoveries in exoplanetary science to come in the next decade, the goal of characterizing a planet to the necessary fidelity to declare it inhabited, not just possibly habitable would appear a very long way off.  Despite the tremendous challenge of this goal, and the pessimism engendered by the current state of our scientific and technical roadmap toward this goal, there is little doubt that this goal is attainable.  The technical challenges, while considerable, are clearly surmountable.  In the four hundred years since Galileo's first use of the telescope, we have achieved more than 20 magnitudes of increased sensitivity, and similarly gains in resolution and wavelength coverage that place us tantalizingly close.  It would seem surprising for this moment to be the one at which our progress would halt.  No doubt the attainment of this goal will be expensive, certainly well beyond any chinks in the astrophysics budgets over the next decade, possibly even the one after that.  That progress will remain indefinitely throttled by short term budgets would seem unlikely.  Life on other planets is a question that fascinates humanity, not just as a scientific question, but throughout popular culture.   It would seem unimaginable to look back from a vantage point another four hundred years into the future and for astronomy to have failed to meet the challenge of identifying how rare or unique life on earth may be.  In this moment when all options appear improbably distant, it is perhaps worth considering new, even if distantly improbable, options.

\section{Identification and Characterization of Biomarkers}

TPF as currently conceived would fall well short of fully satisfying our appetite for knowledge of life elsewhere in the universe.  The presumption has largely been that the appetizer delivered by TPF of the certain knowledge of a habitable planet would lead in several decades to a more ambitious mission, Life Finder.  There has been little imperative to fill out the design as this hypothetical mission has always remained well over the horizon of technology prediction in the domain of speculative futurism.  Here I provide a possible concept that might leap straight to a mission that would offer the capability to both discover earth-like worlds and determine with certainty the status of biological activity on that world.

The identification of "biosignatures", atmospheric constituents that are the signposts of ongoing biological activity remains a subject of intense research.  The principal biologically produced gases on the Earth, O$_2$, its UV-photolysed byproduct O$_3$, CH$_4$, CO$_2$ are all present in at least trace quantities on Mars, where they are predominantly produced abiotically, but CH$_4$ may also be the result of global biological activity \citep{Mumma:2009vn}.  No one gas will constitute a convincing biomarker, but it is the simultaneous presence of gases in large thermochemical disequilibrium that would be the generic sign of a life-bearing world.

Convincing demonstration of the signature of life in an atmosphere will therefore ultimately require not just the detection of bands, but the molecular species and physical state with sufficient detail to account  with reliability for atmospheric and geochemistry to rule out an abiotic origin.  It is hard to believe that the quality of data that we have so far attained on exoplanets would rise to this standard, even if it were available for an Earth-sized planet.

\section{Spectroscopy}

The primary tool by which molecules have been identified in astrophysics and other fields is rotational transitions at microwave frequencies \citep{Townes:1955fk}.  The moment of inertia, and therefore energy associated with the quantum of angular momentum is unique for each molecule.  The ladder of rotational states, splittings and vibrational transitions result in a dense, rich forest of lines, which can be uniquely identified with great certainty. The full molecular inventory of an exoplanet could, in principle, be achieved by observations at very high spectral resolution in the far-infrared/submm.

\begin{figure}[htbp]
\begin{center}
\includegraphics[width=\columnwidth]{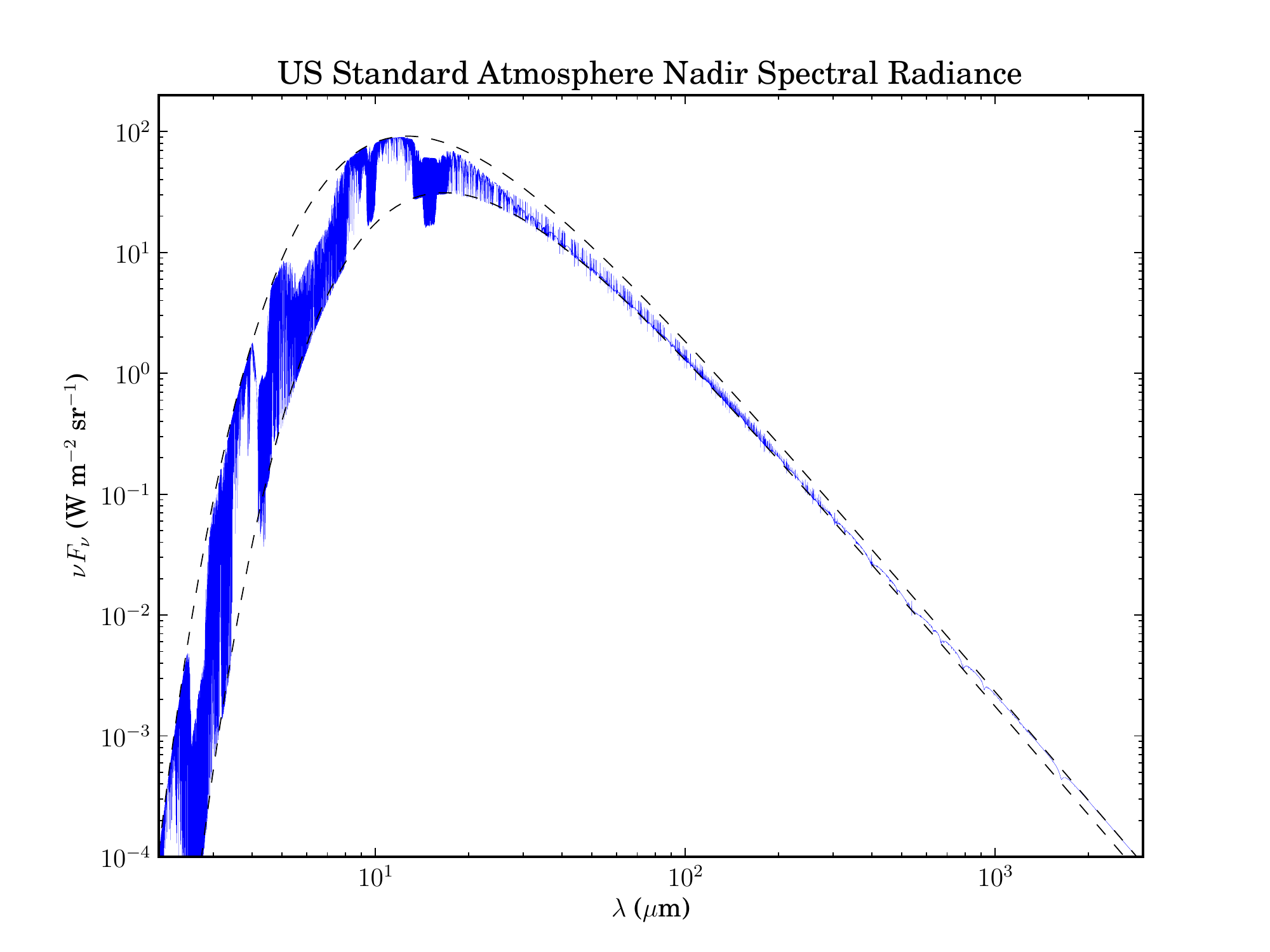}
\caption{Spectral radiance of the US standard atmosphere.  The blue line shows a high spectral resolution calculation of the zenith emergent spectral radiance using the LINEPAK line-by-line code \citep{Gordley:1994fr}.  The calculation is for nadir viewing geometry from an altitude of 600 km with a US standard atmosphere profile and the HITRAN 2008 line opacity data for H$_2$O, CO$_2$, O$_3$, N$_2$O, CO and CH$_4$ and a 288K blackbody for the ground.  The dashed lines show unit emissivity at T=288K and T=220K.}
\label{fig:earthradiance-full}
\end{center}
\end{figure}

The emergent radiance spectrum of the earth (see Figure~\ref{fig:earthradiance-full}) shows a dense forest of absorption features at all frequencies except those at which the atmosphere is completely transparent.  At moderate spectral resolution these features are blurred into bands. By virtue of the higher spectral resolution that is readily achieved at low frequencies, it is predominantly the microwave frequencies at which the most detailed knowledge of the atmosphere is recovered.  

Microwave radiometers, such as the Microwave Limb Sounder on the Upper Atmosphere Research Satellite \citep{Barath:1993kx} are a primary tool in earth observing to determine atmospheric temperature and composition.  Beyond simply detecting lines, the radiative transfer and fitting the observed spectral radiance to a model allows retrieval of full vertical profiles.   The study of atmospheres in the solar system with heterodyne spectroscopy has been proposed for future planetary exploration missions missions \citep{Encrenaz:2001fj,Lellouch:2010uq} and has been reviewed by \citet{Encrenaz:1995qy}, particularly with respect to the prospects of forthcoming (at that time) facilities and more recently with regard to ALMA by \citet{Lellouch:2008kx} and Herschel by \citet{Hartogh:2009rt}.  

Such microwave radiometry has been widely used in ground based remote sensing of solar system objects 
\citep{Biver:2002qf,Boissier:2009bh,Cavalie:2009lq,Cavalie:2008dq,Cavalie:2010cr,Clancy:1983oq,Clancy:1992eu,Clancy:1996nx,Clancy:2004qe,Crovisier:2009kl,Crovisier:2009ai,Drahus:2010lr,Encrenaz:1995uq,Encrenaz:2001fk,Fast:2006fj,Hesman:2007ys,Lellouch:1991mz,Lellouch:1991gf,Lellouch:1991qf,Lellouch:1995ve,Lellouch:2005ly,Lellouch:2008ul,Marten:1993pd,Marten:2005bh,Moreno:2005dq,Moreno:2009lq,Moullet:2008rr,Moullet:2010cr,Rengel:2008nx,Rosenqvist:1992eu,Sandor:2010oq}), 
and observed solar system objects from 
Odin \citep{Biver:2005ul,Biver:2007pd,Biver:2009ve,Cavalie:2008kx,Lecacheux:2003fr}, 
SWAS \citep{Bensch:2004zr,Bensch:2007tg,Bergin:2000gf,Chiu:2001wd,Gurwell:2000uq,Gurwell:2007yq,Lellouch:2002zr,Neufeld:2000wd}, 
Rosetta \citep{Gulkis:2007qy,Gulkis:2008uq,Gulkis:2010lr}, 
and Herschel/HIFI \citep{de-Val-Borro:2010fj,Hartogh:2010qy,Hartogh:2010vn,Hartogh:2010lr}

\begin{figure}[htbp]
\begin{center}
\includegraphics[width=\columnwidth]{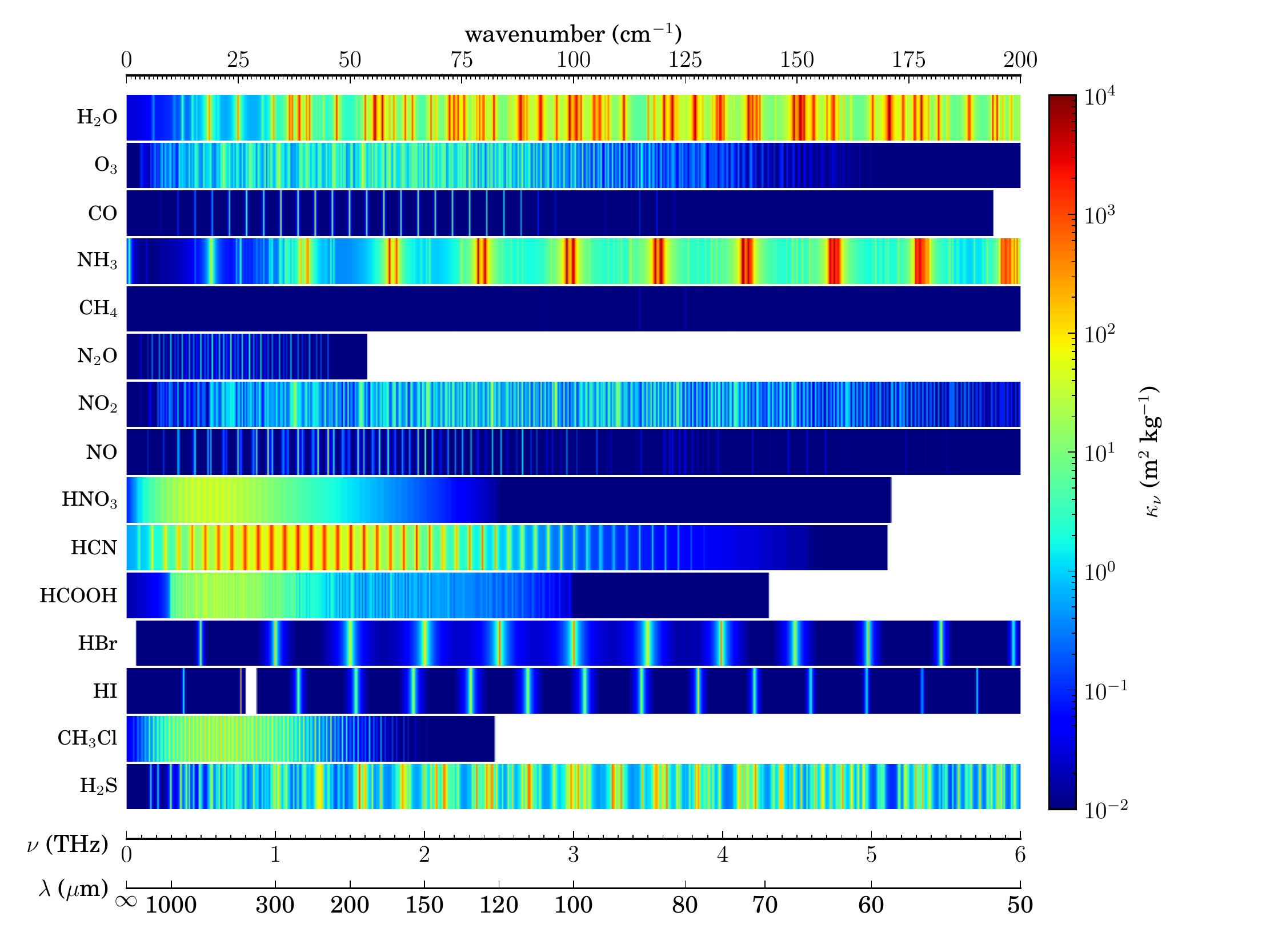}
\caption{Absorption cross-sections in the 50-200 $\mu$m wavelength (1-6 THz) range for molecules of interest in earth-like atmospheres, from the HITRAN 2008 database.  The cross-sections have been broadened for 100 mb and 260K.}
\label{fig:HITRAN}
\end{center}
\end{figure}

\begin{figure}[htbp]
\begin{center}
\includegraphics[width=\columnwidth]{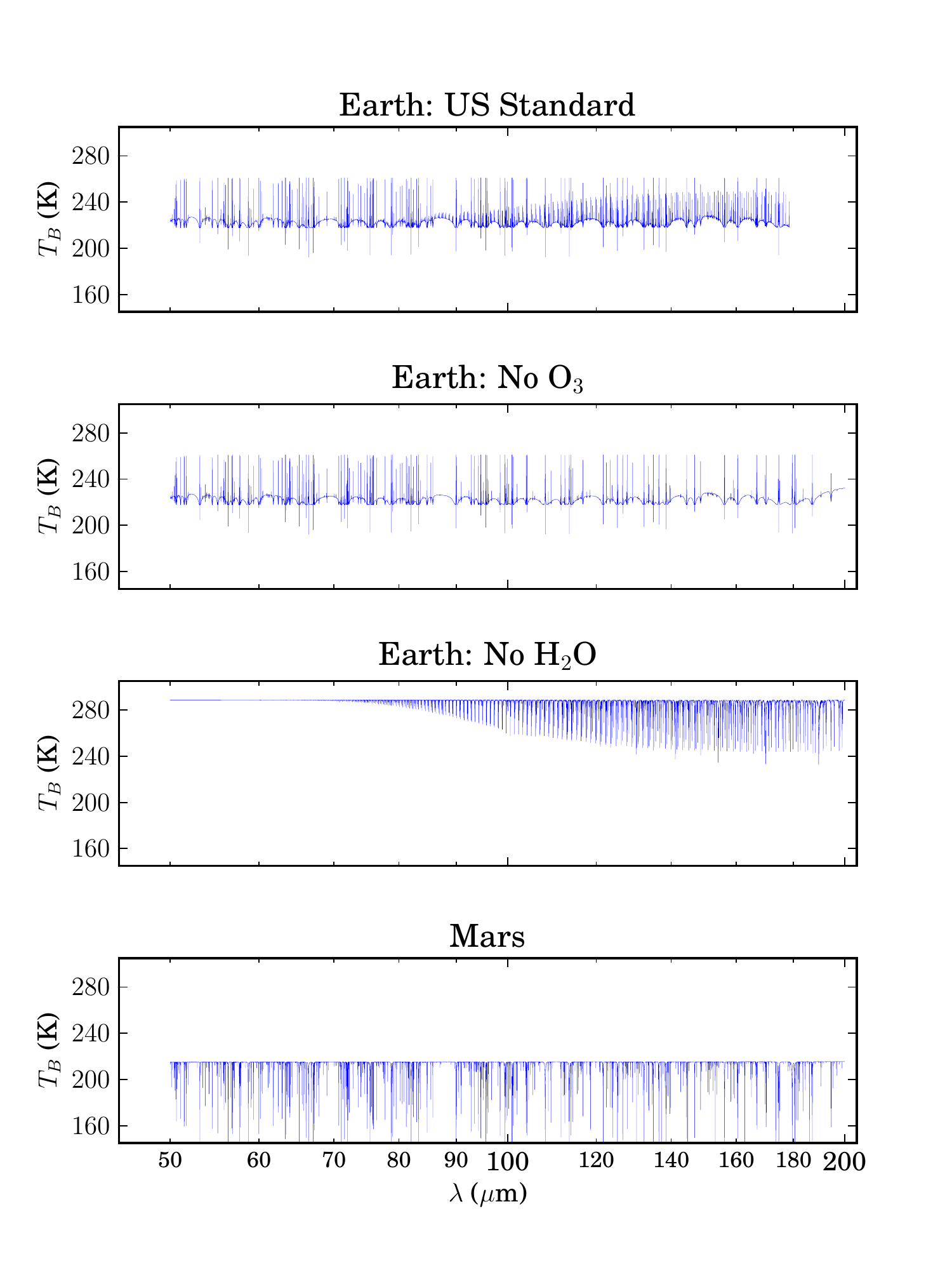}
\caption{Emergent radiance (in brightness temperature) for model atmospheres.   The top panel is the same model as shown in Figure~\ref{fig:earthradiance-full} over the $50-200 \mu$m range, converted to $T_B$.  At wavelengths $< 100 \mu$m the difference between the Rayleigh-Jeans approximation and Planck function becomes significant, so the brightness temperature is numerically derived by solving for the temperature of a unit emissivity Planck function with the equivalent monochromatic spectral radiance.  The second, third and fourth panels show a US standard atmosphere modified by setting the ozone abundance to zero; modified by setting the water abundance to zero and a mars model atmosphere.   The dramatic differences in these spectra demonstrates the sensitivity of far-infrared diagnostics to the molecular constituents of an atmosphere.  }
\label{fig:compare-TbP}
\end{center}
\end{figure}

The wavelength range of $50-1000~\mu$m that is strongly absorbed in the earth's atmosphere is particularly rich in molecular features (see Figure~\ref{fig:HITRAN}) and can provide dramatic diagnostics of even minor variations in atmospheric composition (see Figure~\ref{fig:compare-TbP}).

\section{Coherent interferometry}

The problem of separating the starlight from planet light is fundamentally one of spatial resolution.  To achieve the simultaneous goal of high spatial and spectral resolution immediately suggests the use of coherent interferometry that correlates signals in post-processing rather than interfere light optically.   Temporal correlation of signals measures the spectral content by use of the Wiener-ÐKhinchin theorem, and spatial correlation measures the spatial coherence and therefore source distribution function by application of the van Cittert-Zernike theorem.  This type of interferometry in astronomy is often referred to as heterodyne interferometry, where a heterodyne receiver down-converts the signal to an intermediate frequency, which can be more readily transported by measuring the difference (beat) frequency between the astronomical signal and a local oscillator.   In principle, coherent detection of faint signals can be implemented without down-conversion even at very high frequencies by use of a maser amplifier, although in practice for astronomical purposes heterodyne techniques are more common, and are rapidly being pushed to towards higher operating frequencies and bandwidths \citep{Wild:2010yq}.  

In this paper, we reserve the term heterodyne to refer specifically to the generation of an intermediate frequency with a local oscillator, and use the more general term coherent interferometry whether or not the receiver is heterodyne.  Despite the number of concepts in coronagraphy and nulling interferometry that have been discussed in the literature for terrestrial planet detection, coherent interferometry appears to have been overlooked.  Typically, heterodyne detection is disfavored at short wavelengths, due to the practical limitations of implementable bandwidth, and the fundamental, quantum noise inherent in coherent detectors.  Coherent detection measures the amplitude and phase of the electromagnetic field simultaneously, and so is governed by the Heisenberg uncertainty principle, resulting in a noise equivalent power of one photon per unit bandwidth per unit time \citep{Serber:1960qy}.  In the background (or equivalently system temperature) limit with equivalent bandwidth, direct (total power) detection has an SNR advantage of a factor \citep{Townes:2000fk}

\begin{equation}
\sqrt{ 	\frac{e^{h\nu/kT}-1}{1-\varepsilon}}
\end{equation}

where $h\nu$ is the energy quantum per photon, $k$ is Boltzman's constant, $T$ is the temperature and $\varepsilon$ the transmission of the optics through which the signal is delivered.  For room temperature optics with transmission of 0.9, in the visible, at 0.6 $\mu$m, this factor exceeds $10^{18}$, rendering coherent detection hopelessly insensitive for any astronomical source, although laser heterodyne spectroscopy does have a wide range of laboratory applications in physics and chemistry.  However, at long wavelengths, direct detection does not offer a sensitivity advantage.  For room temperature optics the cross-over is at approximately 100 $\mu$m, although direct detection in practice is often used in astronomy at wavelengths as long as 3mm because of consideration of bandwidth and the availability of large format bolometer arrays.   Coherent detection has a number of distinct advantages (see \citet{Townes:2000fk}), some of which are discussed below.  At wavelengths near 10 $\mu$m, although coherent detection is fundamentally disfavored, the Infrared Spatial Interferometer \citep{Hale:2000uq} has been successfully operating since 1988, and achieving a sensitivity limit of approximately 50 Jy with single-sideband bandwidth of 2.6 GHz.  Contemporaneously, much more ambitious and complex direct detection interferometers have been developed for the Keck and VLT telescopes, which are currently achieving sensitivity limits of
 2.4 Jy \footnote{\url{http://nexsci.caltech.edu/software/KISupport/nulling/index.shtml}} and 
1 Jy \footnote{ \url{http://www.eso.org/sci/facilities/paranal/instruments/midi/doc/VLT-MAN-ESO-15820-3519\_v86.pdf}}.  

Although as expected, the direct detection interferometers are more sensitive, they are fed by telescopes 8-10m in diameter c.f. 1.65m for the ISI, and utilize the full bandwidth of the $8-13~\mu$m $N$ band: 13 THz.   Heterodyne interferometry typically performs very close to the fundamental limits, whereas direct detection interferometry, even though it has a large theoretical advantage, often cannot practically achieve the throughput to fully realize the sensitivity advantage.  Scaling the ISI sensitivity by a collecting area of factor of $(10/1.65)^2$ and increasing the bandwidth to 30 GHz, it would be comparably sensitive to the Keck interferometer.

Since the problem of attaining sufficient sensitivity in the thermal infrared is generally acknowledged {\em not} to be the major challenge for TPF, compared to achieving sufficient resolution and null depth, coherent interferometry is worth considering, even with the anticipated loss of sensitivity.

The limiting sensitivity of a coherent interferometer \citep{Rohlfs:1996kx} is

\begin{equation}
\Delta S_\nu = \frac{ \sqrt{2} k\Tsys }
                                { A_{e1}\sqrt{Nt\Delta\nu} }
\end{equation}

where $A_{e1}$ is the effective area of a single receiver telescope, of $n$ in total, $N=(n(n-1)/2)$ is the number of independent baselines, $t$ is the integration time and $\Delta\nu$ is the integration bandwidth.

Unlike direct detection interferometry, coherent detection allows the signal to be amplified before it is divided for beam combination.  Therefore, the point source sensitivity of the interferometer depends primarily only on the collecting area, not the number of apertures or the beam combination scheme.  For the purposes of sensitivity, we can consider the total effective area of the interferometer, $A_e =  A_{e1}\sqrt{2N}$, which for $n\gg1$ is equal to the sum of the collecting area of the individual elements $n\times A_{e1}$, assuming all baselines are independent and have resolved the source.  This approach conveniently hides all factors related to the number, size and configuration of antennae in a factor (presumed to be) of order unity by assuming that the interferometer is well configured to the problem at hand.

The configuration of an interferometric array is in general a well studied problem in radio astronomy \citep{2001ASSL..268.....K,Cornwell:1988rt}.  The problem of choosing an array design will ultimately boil down to a trade between the size of the array, challenges of formation flying, the precision and complexity of the baseline metrology, and the ultimate impact that the array co-phasing error budget has on the achievable contrast.  
The strongest constraint on the array configuration is that only baselines that resolve the planet from the star, i.e. projected baseline larger than  $\lambda \times (10 \pc / 1 \AU) \sim 200~\m$ at $\lambda/100~\um$ are independent baselines for the purposes of detecting the planet.   The assumption of an effective area that is close to the collecting area necessitates an array on kilometer scales.

On the Rayleigh-Jeans tail, 
\begin{equation}
B_\nu = \frac{2 \nu^2kT}{c^2}
\end{equation}
so for a planet of radius, $r_p$ at distance $d$ ad temperature $T_p$, equating the interferometer sensitivity to the planet flux density, 

\begin{eqnarray}
F_{\nu}  & = & \frac{2\nu^2k T_p}{c^2}\pi\left(\frac{r_p}{d}\right)^2 \\
 &=& 0.3~\mu\mathrm{Jy}
\left(\frac{T_p}{300~\mathrm{K}}\right)
\left(\frac{d}{10~\mathrm{pc}}\right)^{-2}
\left(\frac{r_p}{10^9~\mathrm{cm}}\right)^{2}
\left(\frac{\lambda}{100~\mu\mathrm{m}}\right)^{-2}
\end{eqnarray}

the effective area required is:
\begin{equation}
A_e = \frac{\sqrt{2}}{2\pi} 
\frac{1}{\sqrt{t\Delta\nu}}
\frac{T_{sys}}{T_p}
\left(\frac{d}{r_p}\right)^2
\lambda^2
\end{equation}

For the moment, assuming the ideal quantum noise for coherent receivers, 
\begin{equation}
T_{sys}=\frac{h\nu}{k} = 150~\mathrm{K} \left( \frac{\lambda}{100~\mu\mathrm{m}}\right)^{-1}
\end{equation}

 and that it were possible to implement a coherent receiver with a fractional bandwidth, $\Delta\nu/\nu$, the effective area required is: 

\begin{eqnarray}
A_e & =&  835 \mathrm{~m}^2
\left(\frac{\Delta\nu}{\nu}\right)^{-1/2}
\left(\frac{t}{10^6 \mathrm{~s}}\right)^{-1/2}
\nonumber \\
& & 
\left(\frac{T_p}{300~\mathrm{K}}\right)^{-1} 
\left(\frac{d}{10~\mathrm{pc}}\right)^{2}
\left(\frac{r_p}{10^9~\mathrm{cm}}\right)^{-2}
\left(\frac{\lambda}{100~\mu\mathrm{m}}\right)^{2}
\end{eqnarray}

Although there are optimistic assumptions in this calculation, coherent interferometry typically does perform near theoretical limits, and this figure should be considered as representative of the scale of a realistic mission, although not a specific collecting area.  Noting that the 2.4 m Hubble Space Telescope was approximately 15\% of the collecting area of the largest ground based telescopes at the time of launch, and JWST will be 40\% of the collecting area of the largest current ground based telescopes, comparing to the collecting area of ground based interferometers is suggestive of what could be achieved.  ALMA, which will soon consist of 54 $\times$ 12 m and 12 $\times$ 7 m antennas, is approximately 6500 m$^2$ in total.  The possibility of a space interferometer of this scale does not seem completely out of the question at some point in the future on the appropriate timescale for a habitable planet characterization mission.  

\section{Practicability}

For the purposes of this paper, any concept that appears possible on a timescale of a few decades is worth considering.   Here some of the questions of practicality are considered.  

In comparison to the production of more than 50 12m antennae for ALMA, and the launch of a 6.5m telescope to L2, it shall be supposed that a 10m diameter antenna is feasible for ``mass production'' as a space telescope at some point in the moderately distant future.   At $\Tsys \gtrsim 150~\K$, the telescope cooling requirements are modest, and well within what can be achieved with passive cooling.  Were it possible to achieve the performance assumptions above, 12 such antennae would in principle be sufficient to detect earth-like planets at a rate approaching 1 per day.

The formation flying is significantly less challenging than missions such as TPF-I, TPF-O or LISA, with the most stringent requirements on precision being only on metrology, not closed loop position.  It is advantageous that the antennae need not be positioned near a plane perpendicular to the observing direction (as required for TPF-I) or on the tightly constrained line between source and observer (as required by TPF-O).  Internal delay is more straightforward to implement in a coherent interferometer, and can be as simple as storage memory in the correlator.  Re-pointing the array within the range of internal delays that can be accommodated by the correlator can be achieved solely by re-pointing the individual antennae at the disadvantage that the projected baseline is foreshortened by the a factor of the sine of the angle between  the baseline and source.  An planar array can therefore be pointed within $\pm 60^\circ$ of the array perpendicular with a loss of no more a factor of two in resolution (in one direction only) and no expenditure of formation flying fuel.  

Heterodyne receivers, amplifiers and digital electronics operating at the necessary wavelengths would appear to be the most speculative component.  However, 'Terahertz technology' has recently found broad applications in photonics, materials science, biomedical imaging and even homeland security \citep{Dobroiu:2006lr,Pickwell:2006fk,Wang:2004qy,Fitzgerald:2002uq}, and rapid progress is underway of devices in the appropriate performance envelope.

In the sensitivity estimate above, it is assumed that the full $\Delta\nu/\nu$ bandwidth is integrated for detection.  This would not simultaneously accommodate the recovery of  the high fidelity spectral information as implied above with respect to identification of molecular species.  With sufficient collecting area, it would be possible to measure the planet in narrow bandwidths.  However, in all likelihood the retrieval of composition will not be possible on a line-by-line basis, as it will probably be the case that individual molecular features in fully resolved spectra have SNR $<$ 1.  In principle, template fitting is equivalent to summing the bandwidth of those lines, but the practicality of this challenge will very likely dictate a significant margin in collecting area over and above the assumption above.   Realizing the diagnostic promise of high spectral resolution and at the necessary sensitivity to meet the scientific requirements of life detection will very likely be the most important trade in a practical design.  Sufficient sensitivity at the resolution of the natural linewidths of an earth-like atmosphere $\Delta\nu/\nu<< 1~\km~\persec/c$ does not seem practicable.  However, given that a practicable interferometer could detect planets at a sufficient rate to successfully survey a large number of stars, and the number of planets afforded such detailed follow-up need not be large, this question is best deferred to a trade study in the context of a realistic design.

Very likely a limiting factor will be the computational requirements of correlation, which is orders of magnitude larger than anything so-far conceived or implemented in spectral bandwidth and resolution.  It may be that there are efficient architectures for correlation that are feasible, or that it's necessary to rely on Moore's law, which for extrapolation of 18 months per doubling is a factor of 100 per decade.  In the absence of a specific architecture, which is not considered here, it is not possible to estimate how challenging or when it might be possible to implement a suitable correlator, but it is not a fundamental limitation.

\section{Fundamental Contrast Limitations and Noise Sources}

\subsection{Stellar Thermal Noise}

The fundamental limit the star introduces to the sensitivity of the interferometer to a planet signal is the thermal noise that the stellar radiation adds to the system temperature.  For a star of temperature $T_\star$, radius $r_\star$ and distance $d$, observed with an antenna beam solid angle $\Omega_b$, this thermal noise is 

\begin{equation}
T= T_\star \frac{ \pi(r_\star/d)^2}{\Omega_b} \ \simeq T_\star  \frac{ (r_\star/d)^2}{(\lambda/D)^2} 
\end{equation}

which evaluates to 0.4 mK for $\lambda$=100$~\mu$m, D=12 m and a solar type star at 10 pc.  A coherent interferometer intrinsically separates the stellar signal from the planet signal, because the measurement is one of coherence, not intensity (for an equivalent example in the optical, see \citet{Guyon:2004fr}).  This thermal noise always exceeds the coherent detection quantum noise unless the beam dilution factor times the stellar temperature is of order $h\nu/k$, i.e. the star is close to being resolved by an individual antenna and the quantum noise temperature is less than the stellar temperature.  For $\lambda < hc/kT_\star \sim 2.4 \mu\mathrm{m}$, the quantum noise always exceeds the thermal noise from a solar type star, regardless of antenna size.  Once the antenna fully resolves the star the increased collecting area is offset by the smaller solid angle and the antenna temperature never exceeds the stellar temperature.  For longer wavelengths, the quantum noise always exceeds the thermal noise for $D < \lambda \times (d/r_\star)$.   Therefore, for antennae of aperture diameter less than 1000 m, this noise source is negligible at all frequencies for solar type stars.

This noise source is analogous to the fundamental noise from stellar leakage in nulling \citep{Angel:1997qy}, and it's analogy in coronagraphy \citep{Kuchner:2002lr}.  However, unlike the case for interferometry and coronagraphy, the regime where this noise source dominates is not the regime of practical interest. 

\subsection{Local Zodiacal Background}

At least away from bright galactic emission, the solar system zodiacal cloud is the limiting background radiation noise source.  Approximately, the zodiacal emission can be considered as a uniform thermal emission source at $T=300~\K$ and emissivity $\sim 10^{-6}$, and contributes $300~\uK$ to the system temperature on the Rayleigh-Jeans tail.  For a coherent receiver, this is less than the receiver noise temperature for $\lambda < 50~\m$.  Therefore the thermal noise from the zodiacal background can also be neglected.  

The zodiacal background does show some non-uniform structure as a result of material streaming from comets and recent  collisions. Non-uniformity of this background could result in spurious sources.  The uniform surface brightness at $\lambda=100~\mu$m is $\sim 1~\MJy~\persr \approx 23~\uJy~\persqarcsec$, and so it would require fluctuations in the optical depth of the zodiacal cloud $\delta\tau/\tau$ of order unity on scales of 0.1 arcsec to create a source of comparable brightness to a planet.  It would seem unlikely that such large contrasts or small scales could be present in the zodiacal cloud due to Keplerian shear and particle random velocities.  Further, since the zodiacal cloud is local, the view through the cloud is changing as the earth and could mutually orbit, with an angular speed of order $360^\circ$/yr, or  0.04 arcsec/s.   Therefore, any fluctuation in the zodiacal background will be quickly smeared out.

\subsection{Phase Noise}

The contrast problem for a coherent interferometer is fundamentally different than for coronagraphs or direct detection interferometers.  Heterodyne detection is intrinsically extremely high dynamic range, and has been demonstrated at more than $10^{15}$\citep{Snyder:1988fj}.
Since each element in the interferometer is single-mode coupled, phase errors other than the piston over an antenna propagate only as a gain.  Since all combination is pair-wise, phases are measured uniquely on every baseline, and phase errors can be rejected by closure phase.  Image reconstruction has been achieved at dynamic range exceeding $10^6$ in radio interferometry \citep{Bruyn:2006qy,Bhatnagar:2009yq}, even in the presence of much larger phase errors than would be expected for a space interferometer.   Dynamic range can also be which improved by the more highly constrained fitting of a model to the data analogous to the rejection of speckle noise by non-redundant aperture masking \citep{Lloyd:2006lr}.  In essence, because phase is detected {\em before} any interference, the problem of co-phasing to control scattered light requires only knowledge, not control of the phasing of the array.  The fundamental challenge of wavefront control can be tackled {\em ex post facto} and the knowledge derived from the data itself.   For the purposes of this paper, phasing of the antennae does not constitute a fundamental limitation, although it may well be a practical challenge for formation flying.

\subsection{Sidelobes}

In radio astronomy, the prospect of very bright sources from sidelobes of the primary beam is frequently a major concern.  This is unlikely to be a serious problem at these wavelengths since there are few non-thermal emission sources of significant brightness.  The brightest sources in the sky at $100\um$, Mars, Jupiter and the Galactic center are all thermal emitters of $\sim 10^5 \Jy$.  A perfect $10~\m$ aperture at $100~\um$ has a sensitivity of $\sim 100~\Jy~\perK$, so any source as bright as $10^5~\Jy$ must be at $T_B \ge 1000~\K$ or be fully resolved on baselines larger much larger than the aperture.   Since nearly all bright sources at these wavelengths are cool, if they are bright the cannot be spatially coherent to the interferometer.  Such sources would only contribute thermal noise, which for $\Tsys=h\nu/k\simeq150~\K$ should be negligible if modest care is taken.

\subsection{Exozodiacal Emission}

Thermal emission or reflected light from interplanetary dust in an extrasolar planetary system can be a major complication for planet detection with TPF-I, TPF-C or TPF-O.  In addition to the shot noise from the increased background in the vicinity of the planet, the dust itself may well have asymmetries that are not readily distinguishable from planets.  Dust in the zodiacal cloud is trapped in resonant orbits with the Earth resulting in a 'dust wake' \citep{Dermott:1994fk}, which could be comparably bright to the Earth when viewed as an an unresolved thermal emitter.  To resolve a dust wake requires not just the $\sim$ 1 AU effective (inner working angle) resolution, but significantly less than 0.1 AU resolution, depending on the size of the wake.    In the case of coronagraphs and occulters, which can typically operate with an inner working angle of $<3~\lambda/D$, the requirement of resolving such a dust wake increases the required telescope diameter by an uncertain factor larger than 3.  For an occulter, the occulter also must be commensurately larger and more distant.   Both coronagraphs and nulling interferometers operating at higher resolution would suffer from additional stellar leakage, requiring higher order coronagraphs, or nullers.  For nulling interferometers or band-limited coronagraphs, this resolution limit is fundamental \citep{Guyon:2007qy}.  For this reason, the 2000 and 2010 decadal surveys, and the 2008 Exoplanet Task Force report all require measurements of exozodi to $<$ 10 zodi as a precursor to TPF.  
 
Unlike TPF-I, TPF-C and TPF-O, exozodi does not constitute a risk to successfully detecting a planet for a coherent interferometer.   Cold dust will be well characterized at $\sim 100~\um$ by currently ongoing surveys with Herschel \citep{Eiroa:2010yq,Matthews:2010vn}, that have already demonstrated sensitivity a the level of $L_{\mathrm dust}/L_\star \sim 5 \times 10^{-7}$, comparable to the Kuiper belt, and so the requirement of prior reconnaissance of target stars is already well under way.  Cold dust is at much larger orbital radius than the habitable zone, and will be resolved.   The prospect of characterizing the outer regions of the planetary system is additionally compelling for placing habitable planets in the broader context of planetary system formation and evolution.

\subsection{Background Galaxies}

Despite the major technical challenges of observing in this wavelength regime, the far IR/submm regime has been very rich in astrophysical discoveries for reasons including the specificity and sensitivity to molecular emission which is a driver here, but also because a large fraction of the star formation and AGN luminosity in the universe is reprocessed through dust  and forbidden cooling lines.  Galaxies, particularly ones with active star formation or nuclear black holes, are such prodigious emitters that even ground based far IR telescopes are typically confusion limited \citep{Blain:1998uq}.  At 
$\lambda=100~\um$, the confusion limit for a 10 m telescope is $1~\uJy$ \citep{Benford:2004fj} and rises rapidly at longer wavelengths.  It should therefore be expected that galaxies at cosmological distances will appear with comparable brightness to the postulated earth-like planet within arcseconds of every target star.  These galaxies will, however, be very well resolved by the interferometer, and recognizable by their spectrum.  Features from forbidden lines such as rest frame 158 \um [CII], 63 \um [OI], and PAH bands will be identifiable.  Even if there are sources for which such features are not identifiable, the dust temperature will be sufficiently cold as to be distinguishable from a planet.  The large proper motion of nearby stars will also clearly separate a planet from a background source in $<$ 1 year.  Therefore, although Earth-like planets will be 'confused' with galaxies in the sense that there will be a large number of background sources, they are readily distinguished spectrally or through proper motion.  It is also worth noting that the data on the background sources themselves will be an exquisitely deep parallel survey, equivalent to the cosmological and extragalactic surveys with SPECS \citep{Mather:1998yq}, although with narrower fields unless multi-element receivers can be used to recover a comparable spatial multiplex.

\section{Extension to Planet Imager}

A final opportunity of note with a coherent interferometer is there is a clear path towards more ambitious scientific goals such as resolving the surface of the planet.  Coronagraph and nulling interferometer architectures require dramatic changes to accommodate the change in resolution from the scale of resolving the planet from star to resolving the planet surface.  A free flying coherent interferometer is constrained only by the formation flying performance envelope to expand the baseline.  An interferometer will resolve an Earth-radius object at 10 pc on baselines approaching $2000~(\lambda/100~\um)~\km$, a formation flying challenge which is certainly far less challenging than in all respects than the 5 million km LISA baselines.

Once the planet is resolved, the correlated signal is decreased by the visibility amplitude, so either collecting area or integration time in proportion to $V^2$ will be required.  One would presume that it would be acceptable for imaging followup to proceed at a slower rate than the reconnaissance of stars. It therefore seems feasible to imagine a single interferometer architecture that will be capable of detection, atmospheric characterization and imaging of moderate contrast features on nearby Earth-like planets, rather than a succession of sequential missions such as Terrestrial Planet Finder, followed by Life Finder, followed by Planet Imager.

\section{Conclusion}

A coherent interferometer in the far-infrared offers some compelling advantages for extrasolar planet detection.  Coherent interferometry readily overcomes two of the most compromising challenges to other architectures: the star-planet contrast ratio, and confusion with exozodiacal dust clouds.  The physical and chemical state of an atmosphere could be unambiguously determined.  If, as appears likely, there are no single biosignatures, but instead the only robust sign of life is thermochemical disequilibrium, this approach is uniquely compelling amongst the array of possible techniques to characterize exoplanets in the search for life.  Such an interferometer would further be broadly applicable to a wide range of astrophysical problems.

Under optimistic, but physically plausible sensitivity assumptions, an interferometer with sufficient sensitivity is within a performance envelope of what might be achievable on some speculative timescale.  It is therefore worth considering candidate mission architectures and determining realistic capabilities and technology milestones.

To accurately estimate the yield of information from such a mission, models of far-IR emission from extrasolar atmospheres are necessary.  However, prior work on the characterization of habitable atmospheres has focussed on exclusively on the visible and mid-IR wavelength regimes and there is very little available. Future efforts to model extrasolar atmospheres should extend calculations into the far-infrared.  

\section{Acknowledgements}

The author is grateful to Phil Muirhead, Ray Pierrehumbert, David Spergel, John Storey  and Peter Tuthill for conversations that influenced the development of this concept.  The author's understanding of coherent detection signals and noise owes a great debt to the excellent teaching of Don Backer, whose passing is a great loss to the field of radio astronomy.

\bibliographystyle{apj}
\bibliography{almainspace}

\end{document}